\documentstyle[aps,prb,multicol,epsfig]{revtex}
 \draft
 \begin{document}
 \author{J. Sichelschmidt, A. Loidl}
 \address{Experimentalphysik V, Universit\"at Augsburg, 86135 Augsburg, Germany}
 \author{M. Baenitz, C. Geibel, F. Steglich}
 \address{Max Planck Institut f\"ur Chemische Physik fester Stoffe, 01187 Dresden,
          Germany}
 \author{H.H. Otto}
 \address{TU-Clausthal, Institut f\"ur Mineralogie und Mineralische Rohstoffe,
          38678 Clausthal-Zellerfeld, Germany}
 \title{Quasi-one-dimensional spin chains in CuSiO$_3$: an EPR study}
 \date{\today}
 \maketitle
 \begin{abstract}
Temperature dependent EPR studies were performed on a single crystal of
CuSiO$_3$. This recently discovered compound is isostructural with the
spin-Peierls compound CuGeO$_3$. The EPR signals show different characteristics
than those of CuGeO$_3$ and are due to Cu$^{2+}$ spins located along quasi
one-dimensional chains. For $T>8.2$~K the spin susceptibility closely follows
the predictions of a $S=1/2$ one-dimensional Heisenberg antiferromagnet with
$J/k_B=21$~K. Below $T=8.2$~K the spin susceptibility immediately drops to zero
indicating long range magnetic order.
\end{abstract}


\begin{multicols}{2}
 \columnseprule 0pt
 \narrowtext

The linear spin-chain system CuGeO$_3$ is the first and up to now the only
inorganic compound that exhibits a spin-Peierls transition.\cite{1} Regarding
the magnetic properties the partial substitution of Ge by Si was an important
subject in terms of studying frustration effects \cite{2} and the coexistence
of the spin-Peierls state with long-range antiferromagnetic order.\cite{3,4} To
characterize the nature of antiferromagnetic interactions in Si doped and pure
CuGeO$_3$ electron paramagnetic resonance (EPR) of Cu$^{2+}$ spins provided
important results.\cite{5,6,7,8} In pure CuGeO$_3$ the EPR parameters differ
from those of conventional one-dimensional Heisenberg antiferromagnets. The
antisymmetric Dzyaloshinsky-Moriya (DM) exchange interaction was claimed to
explain this difference.\cite{6} In Si doped CuGeO$_3$ coexistence of
spin-Peierls and antiferromagnetic order is reported for Si concentrations
below $\approx 1\%$. For higher Si concentrations (up to 50\%) a long range
antiferromagnetic ground state is observed.\cite{4} However, for $T>15$~K the
temperature dependence of the EPR parameters does not change significantly for
Si-doping concentrations up to 7\%.\cite{7,8} This paper reports the first EPR
results on pure CuSiO$_3$ which are very different from pure and slightly
Si-doped CuGeO$_3$.

The EPR measurements were performed at X-band frequency with a Bruker ELEXSYS
spectrometer. For cooling a continuous-flow helium cryostat (Oxford) was used.
The spectra were taken from a non-oriented single crystal of CuSiO$_3$ at
temperatures between 4~K and 300~K. DC magnetization measurements at low fields
$H\le10$~kOe were carried out on a commercial SQUID magnetometer.\cite{9} The
single crystal of CuSiO$_3$ was synthesized by dehydration of the mineral
dioptase.\cite{10} Reported EPR spectra of CuO \cite{11} did not show up in our
EPR spectra which indicated the high quality of our crystal. Figure 1 shows the
temperature dependence of  the EPR linewidth and EPR $g$ factor  (determined
from the EPR resonance field) of the investigated CuSiO$_3$ crystal. The inset
of Fig. 1 shows a representative EPR spectrum of CuSiO$_3$ at $T=40$~K (solid
line, derivative of the EPR absorbed power). At all temperatures the spectra
could be nicely fitted with a Lorentzian derivative (dashed line in the inset
of Fig. 1).

\begin{figure}[tbp]
 \centerline{\epsfig{file=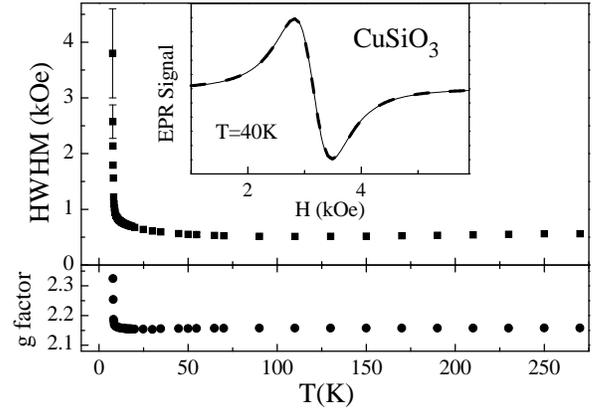,clip,width=8cm}} \vspace{10pt}
 \caption{Temperature dependence of EPR linewidth (HWHM) and EPR $g$ factor,
determined by the resonance field. Inset: typical EPR spectrum (derivative of
absorbed power, solid line) and Lorentzian line fit (dashed line).}
 \label{fig1}
\end{figure}
The EPR linewidth linearly decreases down to $T\approx100$~K at a rate
0.5~Oe/K. This contrasts to a much steeper and non-linear decrease of the
linewidth in CuGeO$_3$ which the antisymmetric DM interaction was used for to
explain the linewidth.\cite{5,6} Therefore in CuSiO$_3$ a DM interaction seems
to be less important. Anisotropic exchange interactions also contribute to the
EPR linewidth. As the deviation of the O(2)-Cu-O(2) angle from $90^\circ$ is
smaller in CuSiO$_3$ than in CuGeO$_3$ \cite{4,10} the anisotropic exchange
interactions in both compounds should be different.\cite{12} For temperatures
above $T\approx12$~K the EPR $g$ factor has a nearly temperature independent
value of $g=2.156\pm0.001$ which is commonly observed for Cu$^{2+}$-ions in an
octahedral environment. This is consistent with the crystal structure of
CuSiO$_3$ which is reported to be the same as in CuGeO$_3$ (orthorhombic, {\em
Pbmm}) and where the Cu$^{2+}$-ions are located within strongly elongated
oxygen octahedra.\cite{10}

\begin{figure}[tbp]
 \centerline{\epsfig{file=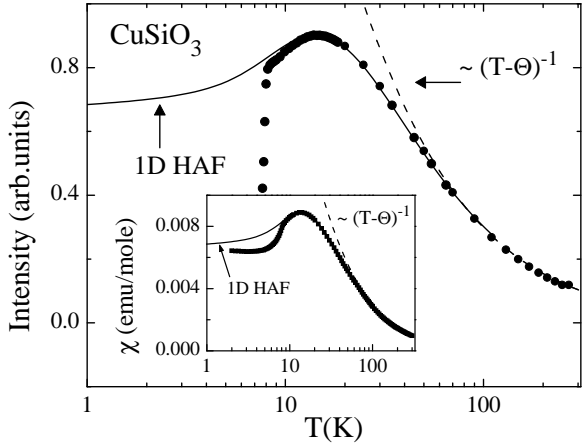,clip,width=8cm}} \vspace{10pt}
 \caption{Temperature dependence of the EPR intensity (integrated EPR signal,
solid circles). Inset: Magnetic susceptibility vs. temperature.\cite{9} The
solid lines represent the behavior of an one-dimensional Heisenberg
antiferromagnet with $S=1/2$.\cite{13} The dashed lines are Curie-Weiss laws
with $\Theta=-7.2$~K.}
 \label{fig2}
\end{figure}

Figure 2 shows the temperature dependence of the EPR intensity $I_{EPR}(T)$
which is determined by integration of the spectra. $I_{EPR}(T)$ is proportional
to the spin susceptibility of  Cu$^{2+}$ and can be well compared with the
magnetic susceptibility $\chi (T)$ \cite{9} as shown in the inset of Fig. 2.
However, below $T = 8.2$4~K the EPR intensity reduces rapidly to zero,
indicating an ordering phenomenon rather than a spin-Peierls state which
produces an exponential decrease of the intensity.\cite{5} This is also
evidenced by the none vanishing $\chi (T\rightarrow0)$ which is usually due to
an anisotropic antiferromagnetic state.\cite{9} Above $T = 8.2$~K $\chi(T)$ and
$I_{EPR}(T)$ are very well described by theoretical calculations for an $S=1/2$
one-dimensional Heisenberg antiferromagnet (1D HAF) without frustration
effects.\cite{13} This leads to an Cu-O(2)-Cu exchange of $J/k_B=21$~K which is
much smaller than in CuGeO$_3$ ($J/k_B\approx160$~K) as can be expected from
the smaller Cu-Cu distances and the smaller O(2)-Cu-O(2) in CuSiO$_3$
.\cite{4,10} For high temperatures $\chi(T)$ and $I_{EPR}(T)$ nicely follow a
Curie-Weiss law with a Weiss-temperature of $\Theta=-7.2$~K, indicating weak
antiferromagnetic coupling.

Figure 3 shows a characterization of the temperature dependence of the EPR
linewidth. The high temperature part is estimated with a linear function
$\Delta H_{lin}(T)=0.5\cdot T\;{\rm Oe/K} + 300$~Oe. This linear part was
subtracted from $\Delta H$ in order to obtain the broadening $\Delta H_{crit}$
when the temperature is lowered towards a critical temperature
$T_{crit}=7.5$~K. A power law $\Delta H_{crit} \propto (T-T_{crit})^{-\alpha}$
approximately describes the linewidth with $\alpha=0.25$ at low temperatures
and above $T=8.2$~K. However, at $T\approx8.2$~K the type of broadening
obviously changes as a noticeable deviation from a power law occurs. This is
indicated by the short dashed line in Fig. 3. The linewidth strongly increases
nearly below the same temperature ($T\approx8.2$~K) where a strong increase of
the $g$ factor is observed as well (see Fig. 1). Hence the change of line
broadening is indicative for the onset of magnetic ordering which yields strong
internal fields and therefore inhomogeneous line broadening effects.
Measurements of the specific heat give strong evidence for long range
antiferromagnetic order.\cite{9} From the critical behavior of the linewidth it
is not possible to compare CuSiO$_3$ unambiguously with typical
antiferromagnets neither for the one-dimensional case like
CuCl$_2\cdot$2NC$_5$H$_5$ ($\alpha=0.5$)\cite{14} nor for the three-dimensional
case like GdB$_6$ ($\alpha=1.5$)\cite{15}.

\begin{figure}[tbp]
 \centerline{\epsfig{file=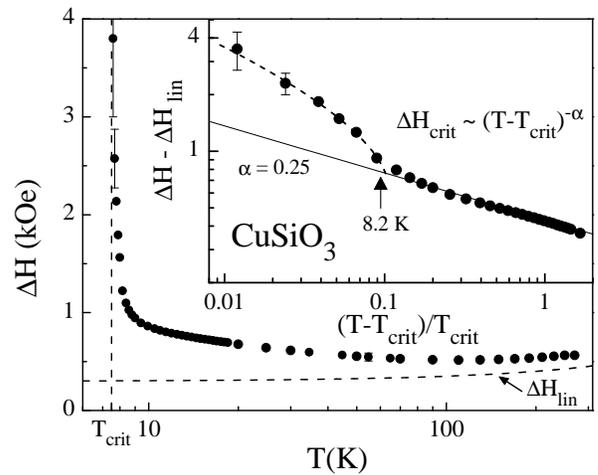,clip,width=8cm}} \vspace{10pt}
 \caption{Temperature behavior of the EPR linewidth $\Delta H$: at high $T$ the
 dashed line approximates the linewidth with a linear function $\Delta H_{lin}$.
 The inset displays the reduced linewidth $\Delta H - \Delta H_{lin}$ vs. reduced
 temperature $(T-T_{crit})/T_{crit}$ ($T_{crit}=7.5$~K) in order to characterize
 the critical line broadening $\Delta H_{crit}$. The solid line represents a
 power law according to $\Delta H_{crit} \propto (T-T_{crit})^{-\alpha}$.}
 \label{fig3}
\end{figure}

In summary our EPR results on CuSiO$_3$ do not show evidences for a
spin-Peierls state below $T=8.2$~K. For low temperatures the EPR intensity and
EPR linewidth are rather explained by long-range magnetic ordering phenomena.
For temperatures above $T=8.2$~K the EPR intensity is proportional to the
magnetic susceptibility and can be reproduced well with a behavior of an
one-dimensional antiferromagnet. Antisymmetric and anisotropic exchange
interactions contribute differently to the EPR parameters in CuSiO$_3$ and
CuGeO$_3$. Further clarification should be provided by measurements at higher
temperatures and at defined crystal orientations which presently are in
progress.
\newpage
We acknowledge fruitful discussions with H.-A. Krug von Nidda and support by
SFB 484.
\end{multicols}

\end{document}